\documentclass[twocolumn,twocolappendix,trackchanges]{aastex63}
\newcommand\dflong{Dragonfly Spectral Line Mapper}
\newcommand\dfshort{DSLM}
\newcommand\nii{\textsc{[Nii]}}
\newcommand\ha{H$\alpha$}
\newcommand\sbunits{$\mathrm{erg} ~\mathrm{cm}^{-2} ~\mathrm{s}^{-1} ~\mathrm{arcsec}^{-2}$}

\usepackage{xcolor}
\usepackage[normalem]{ulem}
\usepackage[caption=false]{subfig} 
\DeclareCaptionFormat{cont}{#1 (cont.)#2#3\par} 
\received{xxx}
\revised{yyy}
\accepted{zzz}
\submitjournal{ApJ}

\shorttitle{A giant shell of ionized gas discovered near M82 with the \dfshort~pathfinder}
\shortauthors{Lokhorst et al.}
\graphicspath{{./}{figures/}}

\begin{document}

\turnoffedit

\title{A giant shell of ionized gas discovered near M82 with the \dflong~\edit1{pathfinder}}



\correspondingauthor{Deborah M. Lokhorst}
\email{deborah.lokhorst@nrc-cnrc.gc.ca}

\author[0000-0002-2406-7344]{Deborah M. Lokhorst}
\affiliation{David A. Dunlap Department of Astronomy \& Astrophysics,
University of Toronto,
50 St. George Street, 
Toronto, ON M5S3H4, Canada}
\affiliation{Dunlap Institute,
University of Toronto,
50 St. George Street, 
Toronto, ON M5S3H4, Canada}
\affiliation{NRC Herzberg Astronomy \& Astrophysics Research Centre,
5071 West Saanich Road, 
Victoria, BC V9E2E7, Canada}

\author[0000-0002-4542-921X]{Roberto Abraham}
\affiliation{David A. Dunlap Department of Astronomy \& Astrophysics,
University of Toronto,
50 St. George Street, 
Toronto, ON M5S3H4, Canada}
\affiliation{Dunlap Institute,
University of Toronto,
50 St. George Street, 
Toronto, ON M5S3H4, Canada}

\author[0000-0002-7075-9931]{Imad Pasha}
\affiliation{Department of Astronomy,
Yale University,
52 Hillhouse Ave., New Haven, CT 06511, USA}

\author[0000-0002-8282-9888]{Pieter van Dokkum}
\affiliation{Department of Astronomy,
Yale University,
52 Hillhouse Ave., New Haven, CT 06511, USA}

\author[0000-0002-4175-3047]{Seery Chen}
\affiliation{David A. Dunlap Department of Astronomy \& Astrophysics,
University of Toronto,
50 St. George Street, 
Toronto, ON M5S3H4, Canada}
\affiliation{Dunlap Institute,
University of Toronto,
50 St. George Street, 
Toronto, ON M5S3H4, Canada}

\author[0000-0001-8367-6265]{Tim Miller}
\affiliation{Department of Astronomy,
Yale University,
52 Hillhouse Ave., New Haven, CT 06511, USA}

\author[0000-0002-1841-2252]{Shany Danieli}\thanks{NASA Hubble Fellow}
\affiliation{Institute for Advanced Study, 1 Einstein Drive, Princeton, NJ 08540, USA}
\affiliation{Department of Astrophysical Sciences, 4 Ivy Lane, Princeton University, Princeton, NJ 08544}

\author[0000-0003-4970-2874]{Johnny Greco}
\affiliation{Center for Cosmology and AstroParticle Physics (CCAPP), The Ohio State University, Columbus, OH 43210, USA}

\author[0000-0001-5310-4186]{Jielai Zhang}
\affiliation{Centre for Astrophysics and Supercomputing, Swinburne University of Technology, Mail Number H29, PO Box 218, 31122 Hawthorn, VIC, Australia}
\affiliation{ARC Centre of Excellence for Gravitational Wave Discovery (OzGrav), Hawthorn, 3122, Australia}

\author[0000-0001-9467-7298]{Allison Merritt}
\affiliation{Max-Planck-Institut für Astronomie, Königstuhl 17, D-69117 Heidelberg, Germany}

\author[0000-0002-1590-8551]{Charlie Conroy}
\affiliation{Harvard-Smithsonian Center for Astrophysics, 60 Garden Street, Cambridge, MA 02138, USA}




\begin{abstract}


We present the discovery of a giant cloud of ionized gas in the field of the starbursting galaxy M82. 
Emission from the cloud is seen in \ha~and \nii$\lambda$6583 in data obtained though a small pathfinder instrument used to test the key ideas that will be implemented in the \dflong, an upcoming ultranarrow-bandpass imaging version of the Dragonfly Telephoto Array. 
The discovered cloud has a shell-like morphology with a linear extent of $0.8^{\circ}$ \edit1{and is positioned $0.6^{\circ}$ northwest of M82. 
At the heliocentric distance of the M81 group, the cloud's extent corresponds to 55 kpc and its projected distance from the nucleus of M82 is 40 kpc. }
The cloud has an average \ha~surface brightness of $2\times10^{-18}$ \sbunits. 
The \nii$\lambda$6583/\ha~line ratio varies from \nii/\ha~$\sim0.2$ to \nii/\ha~$\sim1.0$ across the cloud, with higher values found in its eastern end.
Follow-up spectra obtained with Keck-LRIS confirm the \edit1{existence } of the cloud and yield line ratios of \nii$\lambda$6583/\ha~=~0.340 $\pm$ 0.003 and \textsc{[Sii]$\lambda$6716,6731}/\ha~=~0.64 $\pm$ 0.03 in the cloud.
This giant cloud of material could be lifted from M82 by tidal interactions or by its powerful starburst. Alternatively, it may be gas infalling from the cosmic web, potentially precipitated by the superwinds of M82. Deeper data are needed to test these ideas further. 
The upcoming \dflong~will have 120 lenses, $40\times$ more than in the pathfinder instrument used to obtain the data presented here. 

\end{abstract}


\vspace*{2\baselineskip} 

\section{Introduction} \label{sec:intro}

The M81 (NGC 3031) group of galaxies is one of the richest associations of galaxies in the local Universe. The system has been well-studied at many wavelengths all the way from the gamma ray regime through to the radio. The group contains the nearest ongoing major merger, at 3.66 Mpc \citep[][]{tull13}, with HI observations showing significant neutral gas throughout the field, including clear tidal disruptions and interactions between M81, M82 (NGC 3034), and NGC 3077, the three most prominent galaxies in the group \citep{yun94, chyn08, debl18, sorg19}. Within the complicated debris field of the merger seen at all scales, the individual galaxies are also rich in structure: M81 is a face-on grand design spiral galaxy with a plethora of HII regions \citep{stan14} and star formation extending far past the disk \citep{deme08, okam15, harm17}, while M82 is the nearest starburst galaxy, with large-scale high velocity asymmetrical outflows \citep{shop98}. 

The M81 group is a prime target for mapping diffuse \ha~emission on large scales to search for evidence of circumgalactic gas fueling the star formation in the galaxies. 
Ionized gas visible through \ha~emission is predicted to reside in the circumgalactic medium and halos of galaxies in the local Universe \citep{lokh19}, but is extremely difficult to observe \citep[e.g., requiring stacking of millions of sightlines through galaxies to detect;][]{zhan18}. 
The large scale HI emission encompassing the group \citep{yun94, chyn08, debl18, sorg19} and the well-known ``H$\alpha$ cap'' at a projected 11 kpc distance from the nucleus of M82 \citep{devi99, lehn99} make this group the ideal target for searching for extended ionized emission, which should exist on scales similar to that of the HI emission at very faint levels.

In this paper we present the deepest wide field of view imaging of H$\alpha$ and \textsc{[Nii]} emission from the M81 group published to date. \edit1{These observations were carried out with an upgraded version of the Dragonfly Telephoto Array \citep[Dragonfly;][]{abra14} equipped with}
\edit1{instrumentation to enable ultranarrow-bandpass imaging capability \citep[see][for details]{lokh20}.}
The large field of view and $\approx3''$ resolution of the Dragonfly array, combined with its excellent control of systematics and light scattering, make it well suited to imaging extremely low surface brightness, extended structures, such as ultra diffuse galaxies, galactic outskirts and tidal features \citep[e.g.,][]{merr14, vand15, zhanj18, gilh20}. The addition of ultranarrowband filters enables the detection of low surface brightness line emission from diffuse gas outside the galactic disk \citep[][]{lokh19,blan17}.
In this paper we outline the telescope, observing processes and data reduction procedure for an imaging campaign carried out on the M81 group of galaxies in the Spring of 2020. We present the resulting H$\alpha$ and \textsc{[Nii]} images, and report the discovery of a giant intragroup shell of ionized gas (with a linear extent of $0.8^{\circ}$ or $\approx$ 55 kpc at the distance of the M81 group). Potential origins of the shell are discussed along with our current knowledge of the M81 group.

\section{Observations} \label{sec:observations}

\subsection{Primary \dfshort~Observations}

Narrowband imaging of H$\alpha$ and \textsc{[Nii]} emission from the M81 group of galaxies was collected with a pathfinder version of the \dflong~(\dfshort) located in Mayhill, New Mexico at New Mexico Skies Observatories. The pathfinder \dfshort~is a 3-lens version of the Dragonfly Telephoto Array with Dragonfly Filter-Tilter instrumentation that implements ultranarrow-bandpass imaging capability on the telescope \citep[as described in][]{lokh20}. The pathfinder \dfshort~consists of two Canon 400 mm f/2.8 L IS II USM telephoto lenses and one Canon 400 mm f/2.8 L IS III USM telephoto lens, each with a 14.3 cm diameter aperture. Attached to each lens is an SBIG Aluma 694 camera with a Sony ICX-694ALG CCD sensor, which has an angular scale of 2.45$''$ per pixel, 
resulting in a 1.4$^\circ \times 1.9^\circ$ field of view.
Mounted on the front of each lens is a Dragonfly Filter-Tilter that holds a 152 mm diameter filter. The filters used for these observations have a central wavelength of 659.9 nm and FWHM of 3.1 nm. 
The Filter-Tilters have an allowed filter rotation range of -20$^\circ$ to +20$^\circ$ \edit1{around an axis perpendicular to the optical axis, which enables the central wavelength to be shifted blueward from its intrinsic value by up to 8 nm}. In addition to the Filter-Tilter instrumentation, the Dragonfly pathfinder is equipped with electroluminescent flat field panels which are used to collect flat field images after each science exposure, allowing precise illumination corrections to be obtained at each filter tilt and pointing.

Observations were carried out from February 2020 to May 2020. The observations followed the Dragonfly automated observing model, where the telescope is set up every night for observing at the beginning of the night and the telescope carries out observations autonomously, adapting to changing weather conditions and pausing observations when necessary. In total, this resulted in 73 nights of data collection over the months of February to June 2020 and a total of 652 on-target science frames collected with individual exposure times of 1800 seconds.
The data were taken with the filters at two different tilts: 12.5$^{\circ}$ to target the H$\alpha$~$\lambda6563$ emission line and 7$^{\circ}$ to target the \textsc{[Nii]}~$\lambda6583$ emission line. Tilting the filters smoothly shifts the filter central wavelength, and these two tilts shifted the filter central wavelengths to 656.3 nm and 683.5 nm, respectively. 
The final science images (after removing `bad' frames; see Appendix~\ref{appendix:datared} for details) consisted of a total of 31.7 hr of integration on the H$\alpha$ line 
and 15.3 hr on the \textsc{[Nii]} line with the 3-lens pathfinder \dfshort.
The field of view of the final science images \edit1{is} $\sim2^{\circ}\times3^{\circ}$ after dithering (which was carried out in \edit1{an} 8-point $15'$ pattern) and including the $\sim30'$ offsets of the lens pointings.

\edit1{To verify the existence of some of the features reported here, additional \ha~data were obtained in the Spring of 2021 with the pathfinder \dfshort~on a field located northwest of M82 (targeting a region of bright \ha~emission). The final science image in this pointing consisted of 8.3 hr of integration with the 3-lens pathfinder \dfshort, and was reduced using data processing techniques identical to those used to reduce the initial set of observations, as described below.}

\begin{figure*} [t]
   \centering
   \includegraphics[height=21cm]{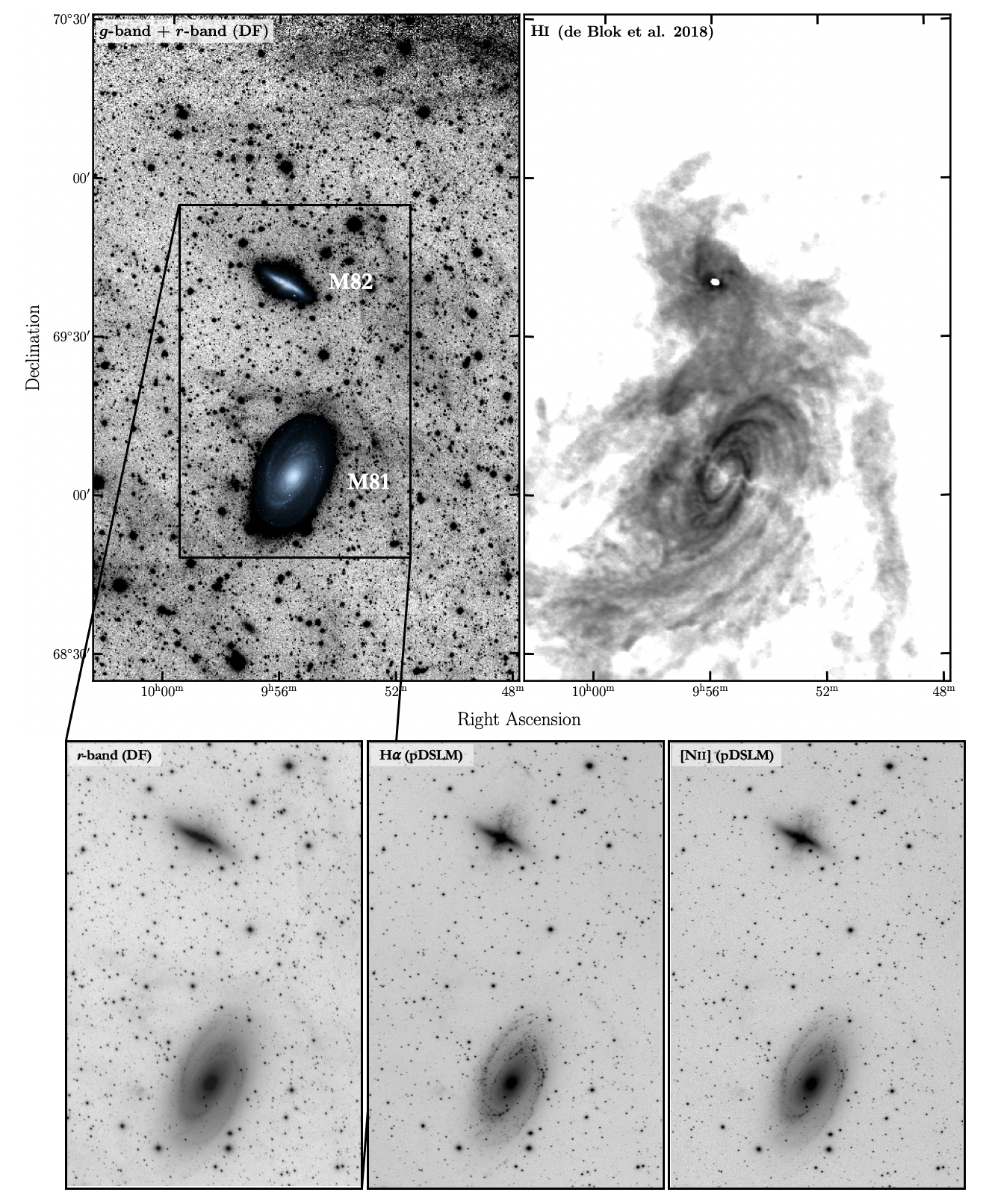}
   \caption
   { \label{fig:contHI} 
    The top left panel displays the Dragonfly Telephoto Array $g$- and $r$-band combined image of M81 and M82 with bright inner galactic regions replaced by continuum-band images from the Digital Sky Survey for reference. This field contains a large amount of galactic cirrus, e.g., the northwest (top right) corner of this field is filled with particularly bright cirrus.
    The HI emission map from \citet[][]{debl18} over the same field of view and log-scaled is displayed for comparison in the top right panel.
    The bottom row displays insets of the above field of view in $r$-band (Dragonfly), \ha~(pathfinder \dfshort), and \nii~(pathfinder \dfshort); these images are log-scaled to showcase both the bright inner galactic regions and faint diffuse extragalactic emission. 
   }
\end{figure*}

Observations of the M81 group were also carried out with the original Dragonfly Telephoto Array \citep[as described in, e.g.,][]{dani20} equipped with broadband $g$ and $r$ Sloan Digital Sky Survey filters. These data were used to subtract the stellar continuum and Galactic cirrus emission from the H$\alpha$ and \textsc{[Nii]} data. 
Broadband observations of the M81 group were carried out on May 16, 2020, gathering a total of 12.5 minutes of integration time on target in the $r$-band and 10 minutes in the $g$-band. Similar imaging was obtained in Spring 2021 for the M82 field pointing.  
The continuum Dragonfly image is shown in the top left panel of Fig.~\ref{fig:contHI}, with the bright inner regions of M81 and M82 replaced by colour composite images from the Digital Sky Survey. The top right panel of Fig.~\ref{fig:contHI} shows an HI emission map of the M81 group of galaxies from \citet[][]{debl18} over the same field of view. The bottom row of Fig.~\ref{fig:contHI} contains smaller field of view insets from the Dragonfly $r$-band image, the pathfinder \dfshort~\ha~image, and the pathfinder \dfshort~\nii~image. 

\begin{figure*} [ht]
\centering
  \includegraphics[height=12.5cm]{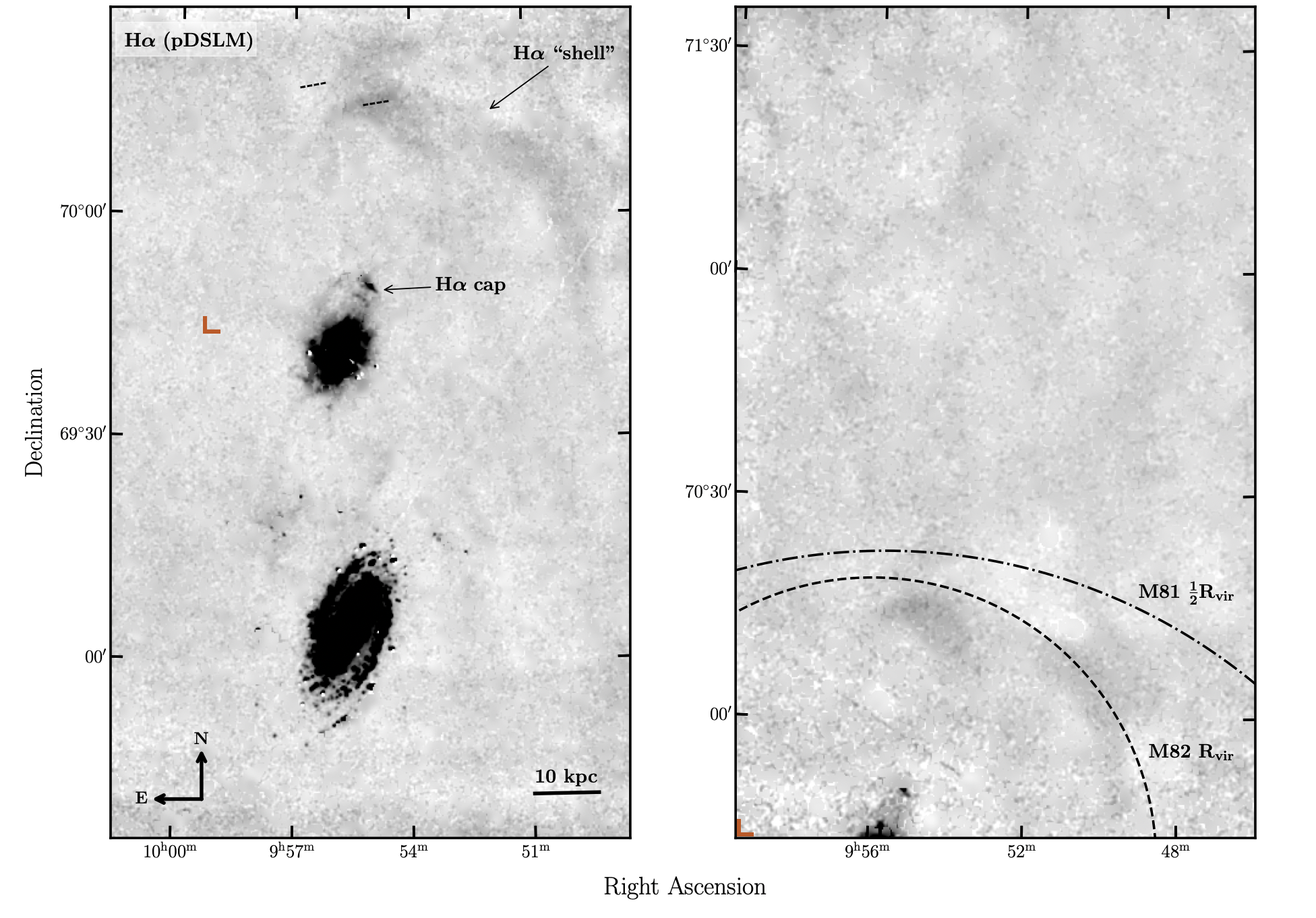} 
   \caption
   { \label{fig:halphaboth} 
   The pathfinder \dfshort~H$\alpha$ map of the galaxies M81 and M82, continuum subtracted, smoothed with a Gaussian filter of 1.5 pixels FWHM and masked. Masked regions are replaced with the median of nearby unmasked pixels. The H$\alpha$ ``shell'' is a region of bright emission in the outer halo of M82. The H$\alpha$ cap is a previously discovered ionized gas cloud in the circumgalactic medium of M82 \citep{devi99,lehn99}. The dashed lines indicate the slit locations used in the Keck-LRIS follow-up observations to the \ha~shell (both off-target and on-target slits are shown). 
   The bands of darker/lighter emission across the bottom and top of the \ha~image are regions of that are only in a subset of the frames due to the dither pattern, and have lower signal-to-noise. 
   A pathfinder \dfshort~H$\alpha$ map of a field located to the northeast of M82 is shown on the right (with the same smoothing and masking as the image on the left). Note that this image has four times less exposure time than the left image, with correspondingly lower signal-to-noise. No extensions to the shell of the same signal-to-noise as the shell are apparent in this image, suggesting that the shell is a condensed structure and not part of a larger Galactic stream of gas. The virial radius of M82 ($\approx44$ kpc) is indicated by the dashed black circle. The half-virial radius of the M81 is indicated for comparison (dash-dotted line) where it is assumed that the galaxies are at the same radial heliocentric distance. \edit1{The red corner mark in both images indicates the southeast origin of the M82 field image for reference.}
   }
\end{figure*}

\subsection{\dfshort~Data Reduction}

The Dragonfly data were reduced using a modified version of \texttt{DFReduce} \citep[see][for a description of the Dragonfly data reduction pipeline]{dani20}. We outline the steps of the pipeline and summarize differing procedures for the narrowband data reduction in Appendix \ref{appendix:datared}.
Continuum light was removed from the final science frames by subtracting a scaled $r$-band image from the narrowband frames. The scaling factor was determined by iteration; as the scaling factor was varied, the M82 galactic disk was monitored to ensure that the continuum emission was subtracted completely while not over-subtracting parts of the disk. 
To determine the error introduced by the continuum subtraction, the emission line flux was calculated \edit1{(using the flux calibration described below)} for two scaling factors which resulted in over- and under-subtracting the M82 disk\edit1{, respectively} \edit1{($r$/\ha~$=$11.3 and $r$/\ha~$=$12.3)}. The resulting fractional error \edit1{in regions of interest} was found to be less than 1\%. This low error is due to the fact that \edit1{the regions of interest have} limited overlap with \edit1{galactic stellar light and Galactic cirrus emission so the noise introduced from continuum sources is minimal. The continuum subtracted \ha~data is displayed in Fig.~\ref{fig:halphaboth}. The left panel shows the M81 group image while the right shows the M82 field image}.

\begin{figure} [t]
   \includegraphics[height=8cm]{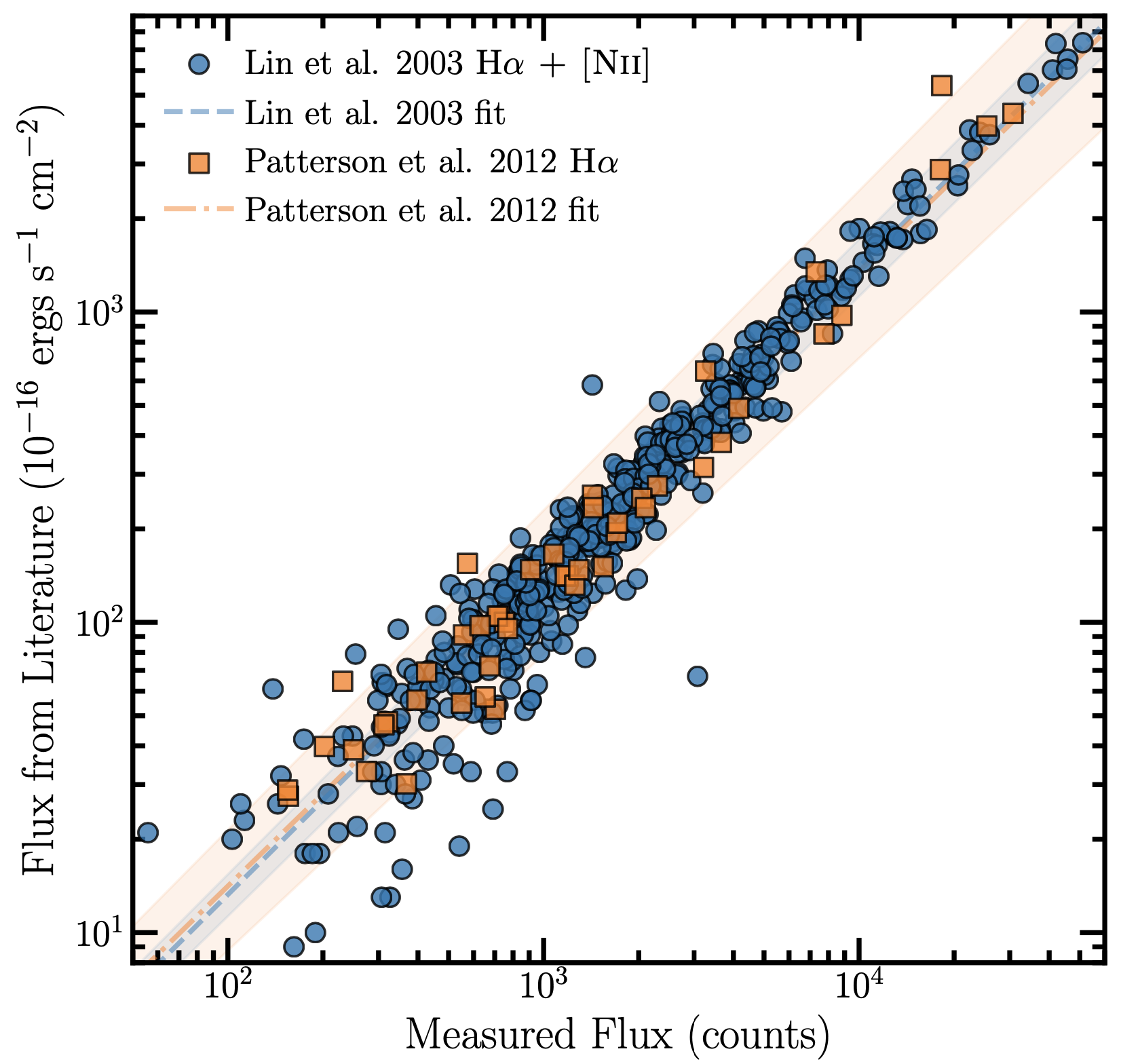}
   \caption
   { \label{fig:fluxcalibration}
   Flux calibration was carried out by comparing the \ha~and \nii~emission flux in M81 HII regions measured in the pathfinder \dfshort~data to published values.
   The blue dashed/orange dash-dotted line corresponds to a linear fit to published fluxes from \citet{lin03}/\citet{patt12}, where the shaded blue/orange region corresponds to 1$\sigma$ uncertainty in the fit determined by bootstrapping and refitting the data to obtain a distribution of fit parameters.
   }
\end{figure}

The images collected by the pathfinder \dfshort~are of such wide field of view that they contain both the M82 and M81 galaxies as well as other members of the M81 group. \edit1{We used the HII regions in M81 to carry out a flux calibration of the images by measuring the emission line flux of HII regions and comparing to values in the published literature on the M81 HII regions.}
Linear fits between the flux in the pathfinder \dfshort~data and published flux values from \citet{lin03} and \citet{patt12} are shown in Fig.~\ref{fig:fluxcalibration}. The two sources of published flux are in good agreement with one another. The resulting fit using both published data sets is $\mathrm{log}_{10}(F~[\mathrm{erg}~\mathrm{s}^{-1}~\mathrm{cm}^{-2}])~=~a~\mathrm{log}_{10}(F~[\mathrm{counts}])~+~b$ where $a=1.00\pm0.01$ and $b=-0.86\pm0.04$. Comparing the fits between the two data sets results in a maximum error of 9\% for the flux calibration. With this calibration, the surface brightness limit of the \ha~data is found to be $\approx5\times10^{-19}$ \sbunits~\edit1{or roughly 0.1 Rayleigh} at the 3$\sigma$ level on a spatial scale of $4'$.

\subsection{Supplementary Keck-LRIS Observations}

\begin{figure} [t]
   \includegraphics[height=6cm]{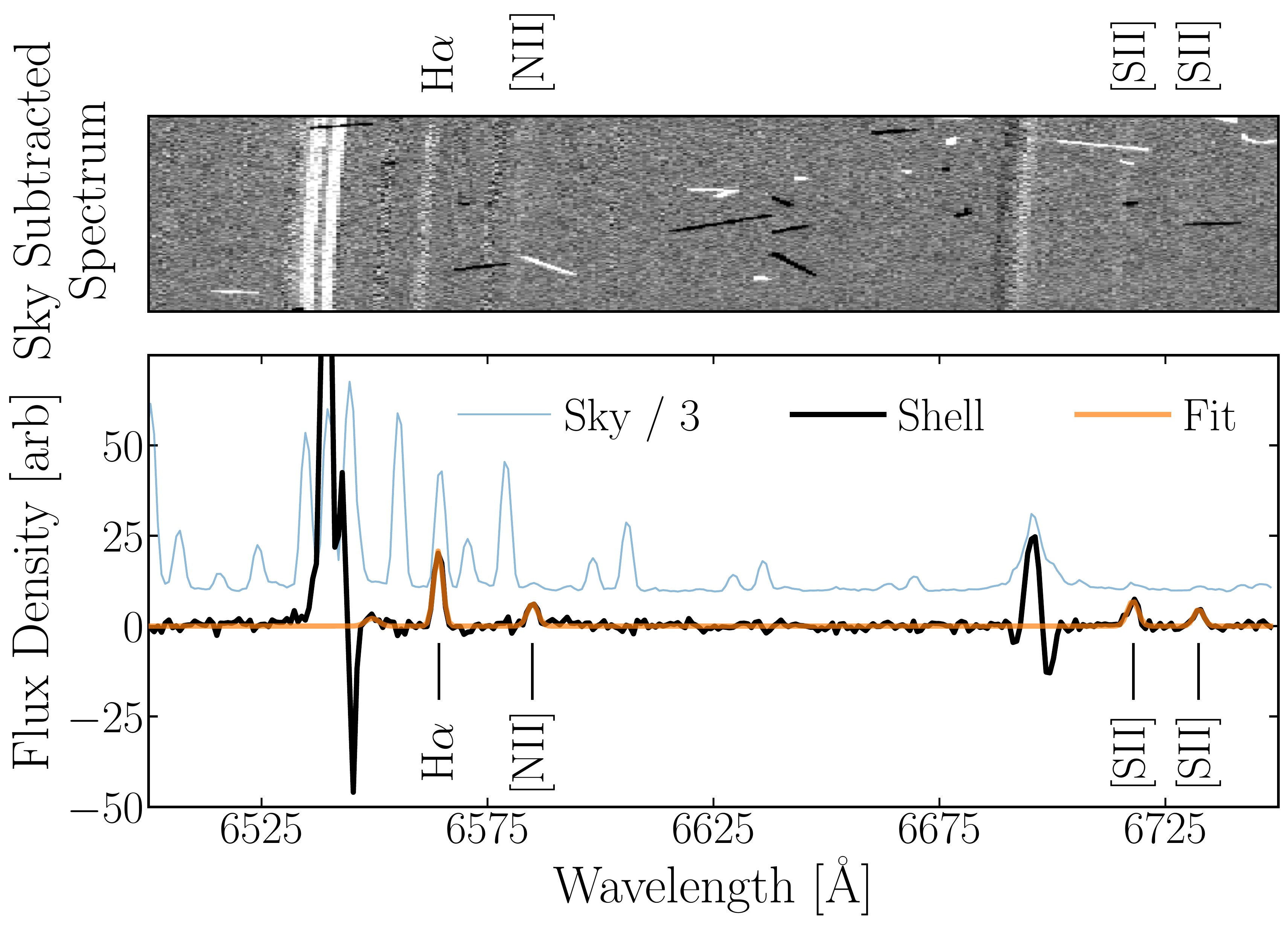}
   \caption
   { \label{fig:LRIS}
    2D (upper panel) and 1D (lower panel) LRIS spectra of the \ha~shell obtained to confirm the existence of the shell and provide basic parameters including radial velocity and relative \nii, \textsc{[Sii]} and \ha~line flux. LRIS spectra obtained at bluer wavelengths (3200 $-$ 5400 \r{A}) show nondetections of the \textsc{[Oii]},
    \textsc{[Oiii]} 
    and H$\beta$ 
    emission lines (not shown).
    }
\end{figure}

Observations with the Keck Observatory Low Resolution Imaging Spectrograph (LRIS) with a $1.5''$ longslit were also collected to verify the \edit1{existence} of observed features in the narrowband data (a total of 1200 seconds on-target and 1200 seconds off-target on both the red and blue sides were collected; the slit locations are shown in the left panel of Fig.~\ref{fig:halphaboth}).  The LRIS spectral data were reduced using \texttt{pypeit} \citep[][]{pypeit:joss_pub} \edit1{and the resulting spectra are shown in Fig.~\ref{fig:LRIS}. }

\begin{figure*}[ht]
    \centering
    \includegraphics[height=13.cm]{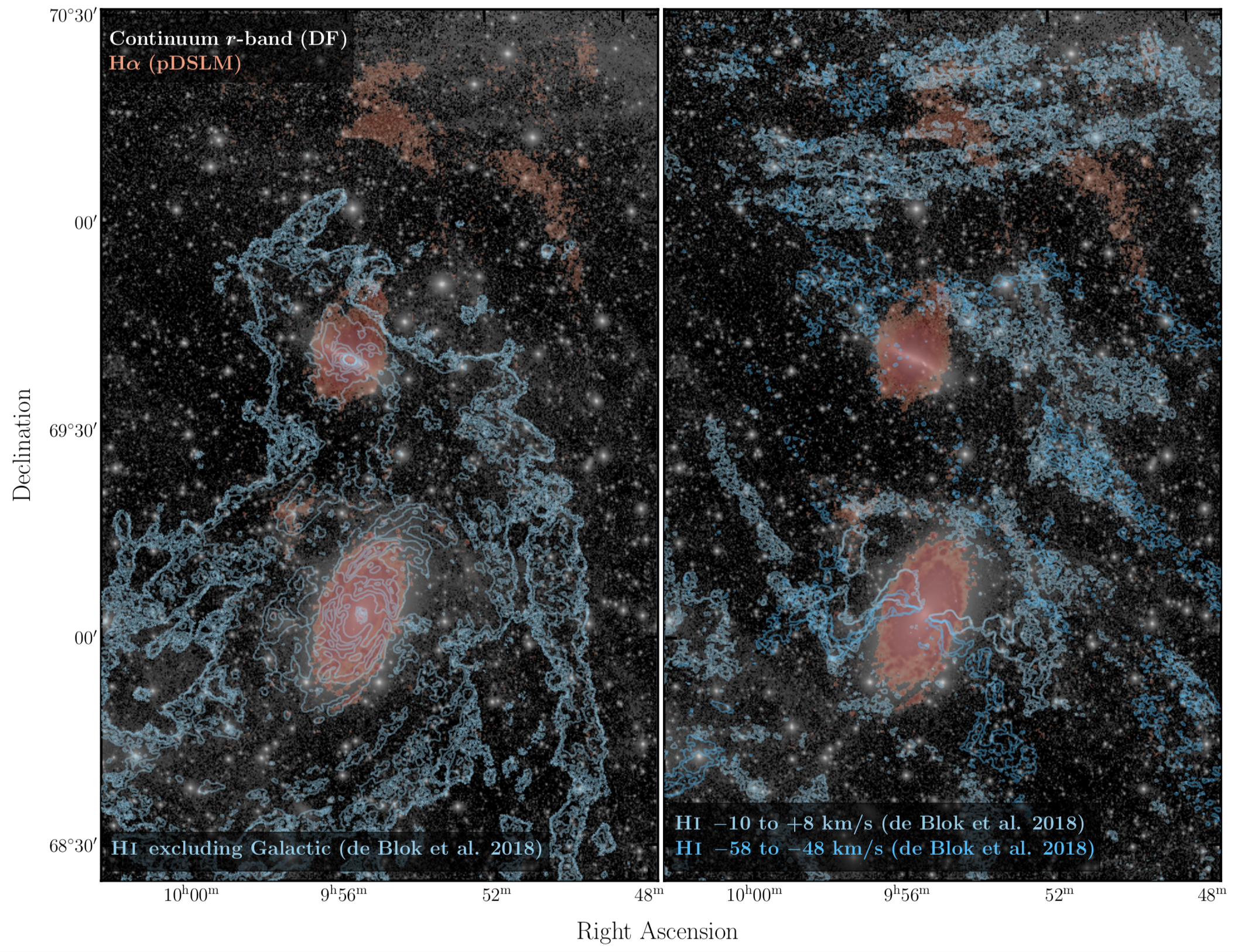} 
    \caption
    { \label{fig:composite} 
    \edit1{A composite image including} high signal-to-noise regions in the pathfinder \dfshort~H$\alpha$ data shown in Fig.~\ref{fig:halphaboth} (red coloured regions) with the $r$-band Dragonfly Telephoto Array continuum image (gray-scale background) and HI emission in the M81 group region \citep[blue contours;][]{debl18}.
    \edit1{On the left, the HI emission is} solely from neutral gas in the M81 group, \edit1{excluding channels containing Galactic HI emission.
    On the right, the HI emission includes only velocity channels containing Galactic HI emission. 
    The HI emission is morphologically distinct from the \ha~shell.}
    }
\end{figure*}

\section{Results} \label{sec:results}

The H$\alpha$ image of the M81 group is displayed in the left panel of Fig.~\ref{fig:halphaboth}. \edit1{This image has been continuum subtracted with bright sources masked and has been smoothed with a Gaussian of 1.5 pixels FWHM.} The \ha~emission traces extended ionized gas in the M81 group of galaxies \edit1{and confirms structures that are already known in the field such as the ``H$\alpha$ cap'' at $\approx11$ kpc from the disk of M82} \citep[][]{devi99,lehn99}. \edit1{It also reveals} additional features that have not previously been detected in \ha. These include a ridge of emission south of the \ha~cap at $\approx8$ kpc from M82 along its minor axis \edit1{which has coincident x-ray emission} \citep[][]{lehn99}. \edit1{Also visible is} an H$\alpha$-emitting filament of gas at the eastern edge of the cap, which is aligned with the HI northern tidal stream \citep[e.g.,][]{debl18}. This feature along with a unique large \ha~emitting clump are discussed in a companion letter \citep[][]{pash21}.

The most striking feature and the focus of this \edit1{paper} is a giant cloud of ionised gas in the field of M82 at the virial radius of the galaxy \citep[$\mathrm{r}_{\mathrm{virial}}~\approx$~44$^{+5}_{-1}$~kpc for a dynamical mass of M82~$\sim$~$1\pm0.4\times10^{10}$ M$_{\odot}$, adopting the R$_{200}$ definition for the virial radius and cosmology with H$_0$~=~70 km s$^{-1}$ Mpc$^{-1}$;][]{grec12}.
The inner edge of the shell of gas is $\sim$~30 kpc from M82 while the outermost edge is $\sim$~46 kpc from M82 in projection.
The  H$\alpha$ shell has an extremely large \edit1{spatial} extent: the projected length of the shell is $\approx$~0.8$^{\circ}$ with an average width of $\approx5.5'$, expanding to $\approx10.5'$ at the shell's widest part, with a total area of $\approx300$ arcmin$^2$. 
At the distance of M82, 3.66 Mpc \citep[][]{tull13}, \edit1{the shell has} a physical size of $\approx$~55 kpc $\times$ $5 - 10$ kpc. The brightest region of the shell has an \ha~surface brightness of 6.5$~\pm~0.3~\times~10^{-18}$ $\mathrm{erg} ~\mathrm{cm}^{-2} ~\mathrm{s}^{-1} ~\mathrm{arcsec}^{-2}$ \edit1{or approximately 1 Rayleigh}. On average, the shell \ha~surface brightness is 1.95$~\pm~0.15~\times~10^{-18}$ $\mathrm{erg} ~\mathrm{cm}^{-2}~\mathrm{s}^{-1}~\mathrm{arcsec}^{-2}$ \edit1{or roughly 0.5 Rayleigh}. The total luminosity of the shell is $L_{\mathrm{shell}}\approx4.7~\times~10^{39}$ $\mathrm{erg}~\mathrm{s}^{-1}$.


Fig.~\ref{fig:composite} displays a comparison of the continuum subtracted H$\alpha$ data (red coloured regions) with the Dragonfly $r$-band continuum image (gray-scale background) and Very Large Array HI data \citep[blue contours;][]{debl18}. 
\edit1{HI emission from the M81 group is shown in the left panel where the velocity channels containing emission from Galactic sources have been removed and are shown separately in the right panel.}
Cirrus emission in the continuum data is clearly visible in the top right corner of the image\edit1{s} (northwest of M82), but the spatial location and physical shape of the cirrus in the broadband image\edit1{s} are distinct from the emission seen in the H$\alpha$ image. There is no significant HI emission associated with the M81 group of galaxies at the location of the M82 shell, though there is one small spatially overlapping region at the southern edge of the shell. In both images, spatially coincident HI and \ha~emission \edit1{have differing morphologies and velocities}, which implies that these gaseous features are not related (this is further discussed below).
The sensitivity limit of the HI map is 1.3$\times10^{19}~\mathrm{cm}^{-2}$ therefore a non-detection at the location of the shell places an upper limit on the neutral gas mass of the shell of $\sim10^{7}$~M$_{\odot}$ \citep[using the quoted HI sensitivity of $\sim$~10$^4$~M$_{\odot}$ per 400 pc resolution element;][]{debl18}.

The \edit1{existence} of the shell was confirmed spectroscopically with independent Keck LRIS observations. 
Fig.~\ref{fig:LRIS} shows spectral data taken along a slit over the shell with the Keck LRIS longslit instrument. The spectra exhibit clear emission lines at the wavelengths of \ha,~\nii~and \textsc{[Sii]}. The emission lines in the spectrum were fit together using the \texttt{Python} packages \texttt{astropy} and \texttt{specutils} to determine a line-of-sight heliocentric velocity of $-$35.4 $\pm$ 4.3 km/s for the shell. The LRIS spectra yield line ratios of \nii$\lambda$6583/\ha~=~0.340 $\pm$ 0.003 and  \textsc{[Sii]$\lambda$6716,6731}/\ha~=~0.64 $\pm$ 0.03, where uncertainties are driven primarily by the sky subtraction.
We estimate the metallicity of the gas (i.e., the oxygen abundance) from the logarithm of the \nii$\lambda6583$/\ha~ratio \citep[the ``N2'' parameter;][]{deni02} using the following equation from \citet{pett04}: 12~+~log(O/H)~=~8.90~+~0.57~$\times$~N2. 
The N2 value derived from the spectral data yields 12~+~log(O/H)~=~8.63, which is about half solar metallicity \citep[0.5 Z$_{\odot}$; using abundances from][]{aspl05}.
However, if the source of ionisation for the shell is shock-based, this metallicity estimate will be artificially skewed to higher values. We will return to the subjects of the ionisation mechanism and accuracy of the metallicity estimate in Section~\ref{sec:ionisation}.

\begin{figure} [t]
   \includegraphics[height=11cm]{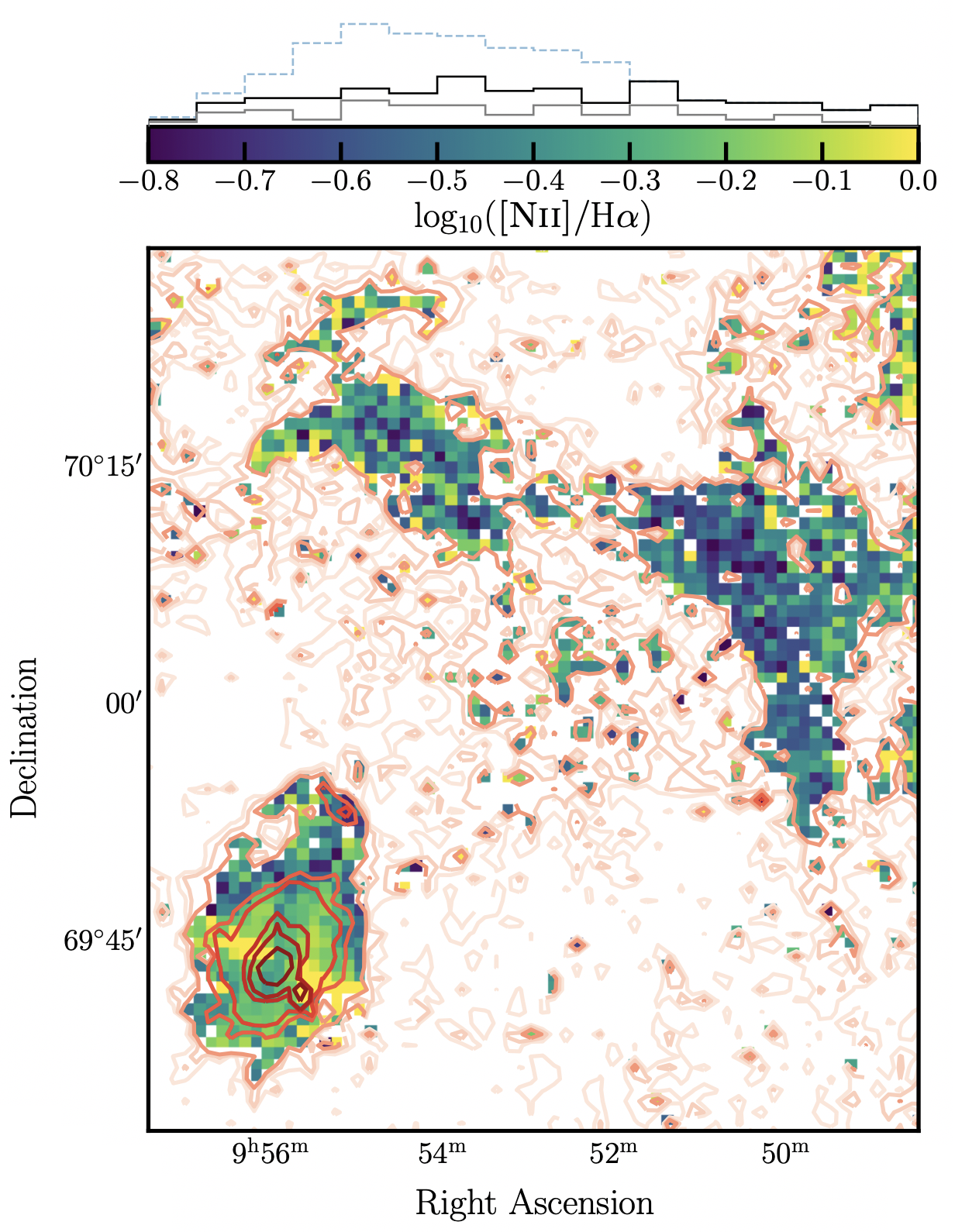}
   \caption
   { \label{fig:figNiiHa}
   Flux ratio map in the vicinity of M82 and the shell of the \nii~$\lambda6583$ emission line flux to the \ha~$\lambda6563$ emission line flux. The map is log scaled between -0.8 and 0. A contour map of \ha~data is superimposed (red shaded contours). Regions of low signal-to-noise in the \ha~emission line are removed from the ratio map, which regions with low signal-to-noise in the \nii~emission line are replaced with the sky background level as upper limits.
   A histogram of the number counts of \nii/\ha~values in the shell is shown above the colour bar. The blue dashed lines include regions that use upper limits for the \nii~values, \edit1{whereas the black and grey solid lines do not include regions that use upper limits for the \nii~values and have applied a 1$\sigma$ and 2$\sigma$ cut on the \ha~data, respectively.}
   }
\end{figure}

A zoom-in comparison of the \nii~$\lambda6583$/\ha~$\lambda6563$ flux ratio in the vicinity of M82 and the shell is shown in Fig.~\ref{fig:figNiiHa}. Overlaid red-shaded contours indicate levels of H$\alpha$ flux in the image. Regions with \ha~emission less than $1\sigma$ above the background level are removed from the ratio map. In the case of \nii, regions with signal-to-noise ratios $<1$ are replaced with the background level, yielding upper limits. The \nii/\ha~ratios detected in the shell range from as high as $\approx1$ along the inner edge of the shell down to \nii/\ha~$\approx0.16$. \edit1{This range is consistent with the \nii/\ha~ratio derived from the Keck LRIS spectra (0.34), which falls close to the middle of the range observed from the narrowband imaging (0.16 $-$ 1). 
At the location of the LRIS slit, the \nii/\ha~ratio from the narrowband imaging is  \nii$\lambda$6583/\ha~=~0.39 $\pm$ 0.09, 
which is consistent with the LRIS ratio within the uncertainties.
The large spread in \nii/\ha~observed in the shell} could be due to differing enrichment of the gas in the shell as a function of location, or due to differing ionisation mechanisms \citep[e.g., shocks and active galactic nuclei produce higher line ratios than HII regions due to the harder photoionizing radiation, e.g.,][]{kewl19}. 

We estimated the density of the shell from the \textsc{[Sii]}~$\lambda$~6717/\edit1{$\lambda$}~6731 line doublet ratio. The ratio between the two line intensities is $1.6\pm0.1$, which places the cloud firmly in the low density regime with an upper limit on the electron density of $n_e \lesssim 1$ cm$^{-3}$ \citep{oste06}. 
Another estimate for the \edit1{average density in the shell} can be made from the \ha~surface brightness measurement, assuming the shell is in thermodynamic equilibrium and case B recombination, with the equation $I_{H\alpha} = 8.7\times10^{-8} E_m$ erg s$^{-1}$ cm$^2$ sr$^{-1}$ \edit1{where $E_m$ is the emission measure} \citep[][]{spit98}. This yields 
\edit1{$n_e\sim 1.4 \times 10^{-4} (L_{s}/5 \mathrm{kpc})^{-1/2} (\eta/10^{-4})^{1/2}$ cm$^{-3}$}, 
where $\eta$ is the volume filling factor of the shell \citep[volume filling factors are typically between $10^{-6}$ and $10^{-1}$ for HII regions;][]{ho97} and $L_{s}$ is the line-of-sight length of the shell.
The total mass of the shell using the $n_e$ estimate from the \ha~surface brightness is 
\edit1{$M_{H} \sim 5 \times 10^6 (L_{s}/5 \mathrm{kpc})^{1/2} (\eta/10^{-4})^{1/2}$ M$_{\odot}$}.
This mass is similar to that of a giant molecular cloud and below the detection limit of the \citet[][]{debl18} HI map.


\section{The origin of the H$\alpha$ shell} \label{sec:discussion}


As discussed earlier, the M81 group of galaxies is an extremely active group of galaxies yielding many possibilities for the origin of the shell. It could be gas lifted from the disk of M82 by the powerful central starburst (as originally argued for the origin of the \ha~cap) or gas falling in onto the galaxy for the first time from the intergalactic medium. 
\edit1{The umbrella-like shape of the cloud} centered on the location of M82 and \edit1{in line} with the minor axis of M82 strongly suggests an origin or excitation mechanism related to the starburst of M82.
The giant shell could be tidally stripped from the group of galaxies similarly to the wide-spread tidal features traced by HI emission, which were produced by close passages between M81 and M82, the last of which was 200-300 Myr ago \citep[e.g.,][]{cott77,yun93,yun94}.
\edit1{The similarities of size and position between the neutral and ionized gas can be seen in Fig.~\ref{fig:composite}}. 
Alternatively, the cloud could be a chance projection of gas in the Galaxy and be unrelated to the M81 group.
We discuss constraints on the origins of the \ha~shell based on the viability of ionisation mechanisms, along with the mass, metallicity and velocity of the shell.

\subsection{Ionisation mechanism of the shell} \label{sec:ionisation}
The \ha~emission from the shell could be produced through photoionisation (e.g., by the local or global ultraviolet ionising background [UVB], active galactic nuclei or starburst ionising radiation or young OB stars) or shock ionisation of the gas. 
The shell could also be a light echo of line emission from a past period of bright emission of a nearby source, such as a peak of starburst emission from M82.
We start by determining the photon ionising flux required to produce the observed level of \ha~emission. We can then compare that to the available photon flux whether the cloud is positioned within the M81 group or within the Galaxy. 
Assuming ionization equilibrium and case B for optical recombination emission, the number of ionising photons, $Q_{H^0}$, for the optical line recombination emission is given by \citep{oste06}:
\begin{equation}\label{eqn:halpharecomb}
        Q_{H^0} = \frac{L_{H\alpha}}{h\nu_{H\alpha}}\frac{\alpha_B}{\alpha^{eff}_{H\alpha}},
\end{equation}
where $L_{H\alpha}$ is the luminosity of the \ha~emission, $\alpha_B$ is the recombination coefficient for case B \ha~emission \citep[2.59$\times10^{-13}$ cm$^3$ s$^{-1}$ at temperature T = 10$^4$ K;][]{oste06}, and $\alpha^{eff}_{H\alpha}$ is the \ha~effective recombination coefficient \citep[11.7$\times10^{-14}$ cm$^3$ s$^{-1}$ at temperature T = 10$^4$ K and electron density n$_e\approx1$ cm$^{-3}$;][]{spit98}. Assuming that the cloud is roughly spherical, we can directly estimate the photoionisation rate and compare it with estimates for the UVB. \edit1{Note that adjustments to this geometry (e.g., cylindrical or planar) result in the same order of magnitude estimate \citep[e.g.,][]{dona95}}. Assuming a power-law approximation for the photoionisation cross-section \citep[][]{oste06} yields the following equation for the photoionisation rate, $\Gamma$:
\begin{equation}\label{eqn:uvb}
        \Gamma=\frac{4\pi A_0}{h}\frac{J_0}{\beta+3},
\end{equation}
where $h$ is Planck's constant, $A_0$ is the photoionisation cross-section at the Lyman limit, $J_0 (\nu)$ is the source of ionizing photons, and $\beta$ is the power law coefficient for $J_0 (\nu)$ \citep[commonly assumed to be 1.8 for the redshift $z=0$ UVB; e.g.,][]{adam11}. 
Together, \edit1{Equations \ref{eqn:halpharecomb} and \ref{eqn:uvb}} yield an estimate for the ionising source of photons to have a photoionisation rate $\Gamma\approx2\times10^{-12}$ s$^{-1}$. This is two orders of magnitude higher than the current estimate for the global UVB at redshift z = 0 \citep[e.g.,][]{fuma17}, indicating that there needs to be an additional source of ionisation for the shell.

If the shell is part of the M81 group, alternate possibilities for ionizing the shell are photoionisation by high energy photons from the starburst in M82 or shock ionisation.
We first consider whether it is viable for the ionising photons from the starburst in M82 to ionise the gas in the shell and produce the observed \ha~emission.
The ionising flux from the M82 starburst is $\sim10^{54}$ photons s$^{-1}$ \citep[][and references therein]{mcle93}, so the intercepted fraction by the shell would be $\approx$~3 -- 23 \% of the ionizing flux, where the line-of-sight width of the shell is assumed to range between the transverse height ($\approx9$ arcmin) and length ($\approx55$ arcmin). Taking the number of \ha~photons from the shell as $\approx1.56\times10^{52}$ photons s$^{-1}$ \edit1{and applying Equation~\ref{eqn:halpharecomb}} estimates the required number of ionizing photons to be $\approx3.5\times10^{52}$ photons s$^{-1}$. 
\edit1{For shell line-of-sight widths ranging from $\approx9 - 55$ arcmin, the required escape fraction is $\sim95 - 20$\%.} This is significantly higher than that \edit1{observed} in the Galaxy or in dwarfs \citep[$\sim$6\% and $\sim$3\%, respectively; e.g. ][]{blan99,zast11,barg12}. 
This implies that photoionisation of the shell from the M82 starburst may be possible only if the superwinds from the M82 starburst cleared the field of M82, or if there is anisotropic radiation from the starburst with flux focused along the minor axis of M82. 
\edit1{The shell emission could be an ``ionisation light echo'' produced by ionising photons from a peak in the M82 starburst that occurred $\gtrsim2\times10^{5}$ years ago and has since dropped off. \citet{fors03} determined that there were peaks in starburst activity $~5-10\times10^6$ years ago, which lasted a few million years and produced at least an order of magnitude greater ionizing flux during its peak than is currently observed. The recombination timescale of the hydrogen atoms for the electron density estimate in the shell of $n_e\sim10^{-4}$ $\mathrm{cm}^{-3}$ is $\sim10^9$ years, so one would expect to still see \ha~emission from ionisation that occurred $\sim10^{7}$ years ago. Accounting for this increase would ease the requirements on the high escape fraction from M82 to photoionise the shell.}

Shocks from the accretion of gas onto the M81 group or M82 halo or shocks produced by the M82 superwinds incident upon the shell could \edit1{also play a large role in} the ionisation of the gas. The disruption of an infalling cloud itself may also produce shock emission \citep[][]{blan07}.
The higher end of the range of \nii/\ha~ratios observed in the shell (\nii/\ha~$\gtrsim0.6$) implies a harder radiation field than that produced by normal star-forming regions \citep[e.g.,][]{kewl19}, \edit1{which suggests} that shock ionisation must play a role in the shell.
The \ha~cap is thought to have been ionised through both slow and fast shocks from the M82 superwinds with shock velocities of $\approx50$ and 800 km/s, respectively; perhaps a similar situation is occurring in the shell \citep[][]{mats12,lehn99}. 
The higher end of \nii/\ha~ratios from the narrowband images (\nii/\ha~$\approx0.3 - 1$) and \textsc{[Sii]}/\ha~ratios from the spectra
are consistent with shocks of velocity between $80-350$ km/s \edit1{into gas of solar (Z$_{\odot}$) to double solar (2~Z$_{\odot}$) metallicity based on the MAPPINGS III fast shock models from \citet[][see their Fig. 21]{alle08} and slow shock models of \citet{rich10} and \citet{raym79}.} 
The lower end of \nii/\ha~values found in the shell (\nii/\ha~$\approx0.16 - 0.4$) are consistent with shocks into gas with metallicity of $\approx0.25$~Z$_{\odot}$ \citep[without further line information the shock velocity in this case cannot be narrowed down;][]{alle08}. 
The line ratios measured in the spectra have low line widths, consistent with velocity dispersion $\lesssim50$ km/s, which suggests that slow shocks rather than fast shocks are responsible for the ionisation of the shell.
The range of \nii/\ha~ratios observed in the shell may be produced by a range of shock velocities across the shell (e.g., a higher shock velocity at the north-eastern edge of the shell where the emission is brightest, which is \edit1{in line} with the superwind outflow from M82).
In order to confirm whether a shock is ionising the gas in the shell and to pinpoint the shock velocities, followup observations are required. For example, deeper spectroscopy targeting the optical or UV emission lines (such as the \nii/\ha~and \textsc{[Oiii]}/H$\beta$ line ratios) will allow better constraints on the required shock velocity. Optical line ratios such as \textsc{[Oiii]}$\lambda5007$/\textsc{[Oiii]}$\lambda4363$ and \textsc{[Nii]}$\lambda6583$/\textsc{[Nii]}$\lambda5755$ and X-ray observations can also be used to determine the temperature of the gas, which is related to the shock speed.

\subsection{Is the shell part of the M81 group or the Galaxy?}
The morphology, size and position of the \ha~shell strongly suggest that the shell is associated with the M82 galaxy, but due to the possibility of a chance projection of Galactic clouds with the M81 group it is worth considering the possibility that the shell is associated with the Milky Way galaxy.
Associations between gaseous features appearing close in projection to galaxies can be made by comparing the radial velocities of the gas to the group and its members \citep[e.g., see analogous arguments made by][in their discovery of a gas cloud near M51]{watk18}, but the M81 group of galaxies is near rest velocity with respect to the Milky Way.
\edit1{The velocity of the cloud is not enough to differentiate between an association with the Galaxy or the M81 group.}
The radial velocities of M81 and M82 are $\approx-15$ km/s and $\approx210$ km/s, respectively \citep[e.g.][]{yun94,beck78,mcke93}, while the velocity of the shell is measured to be $\approx-35$ km/s from the LRIS spectra ($\approx-250$ km/s with respect to M82). The M82 \ha~cap contains gas at velocities between $-$250 to 250 km/s with respect to the systematic velocity of M82 \citep{devi99} while the radial velocity of the HI content in the M81 group ranges between $-270$ to 320 km/s \citep[e.g.,][]{debl18}. The radial velocity of the shell is consistent with both of these velocity spreads. 

\edit1{No extensions to the shell of equivalent signal-to-noise are apparent in the \ha~pathfinder \dfshort~imaging shown in the right panel of Fig.~\ref{fig:halphaboth}, suggesting that the shell is a condensed structure and not part of a larger Galactic stream of gas.}
\edit1{High latitude clouds in the Milky Way, such as high velocity clouds or HVCs, are defined by their neutral gas content.} There is a lack of HI in the region of the shell which is unusual for HVCs. The nearest HVCs to the M81 field are Complex C and Complex A, which skirt the M81 group region. The HVCs and the H$\alpha$ shell are spatially separated in projection by an angular distance of more than 4$^{\circ}$ \citep{west18}. In addition, the velocities measured from \ha~emission and HI in the A and C Complexes are between $-$165 to $-$180 km/s and $-$111 $\pm$ 2 km/s, respectively \citep{tuft98}. This is significantly different \edit1{at the 10$\sigma$ level} from the measured \ha~velocity of the shell (-35.4 $\pm$ 4.3 km/s). However, catalogues of known HVCs are not necessarily comprehensive, since emission sources with velocities within the ``deviation velocity'' (e.g., $\pm50$ km/s about the velocity of the Galactic disk) are often not included when searching for HVCs.
We have inspected this velocity gap using the \citet{debl18} HI emission measurements covering the M81 group, which contain channel maps between $-$58 to $-$48 km/s and $-$10 to +8 km/s that are associated with Galactic emission (all other channels are determined to be associated with the M81 group). \edit1{The HI emission within those channels is compared to the \ha~emission in the right panel of Fig.~\ref{fig:composite}, in which the HI and \ha~data are plotted as blue contours and red shading, respectively. The  morphology  of  the  HI  gas  at  these  velocities  is  distinct  from  the  shell,  with  filamentary HI emission spanning the field-of view.  Spatially coincident HI emission does not appear to be coherent with the \ha~shell, differing in both shape and size.}

\edit1{Another indicator of whether the shell could be part of the Galaxy is if there is a viable ionisation mechanism that could produce the observed emission.} If the shell is in the Galaxy, the expected photoionisation mechanism would be either young OB stars (such as in HII regions) or the local UV background. In Section~\ref{sec:ionisation}, we showed that the global UVB would not be enough to ionise the cloud such that it produces the observed \ha~emission surface brightness. Additionally, we see no evidence for bright ionising stars in the broadband data, or for ionization from localized sources within the cloud -- the cloud appears to be uniformly illuminated at the resolution scale of our data. 
The image pixel scale corresponds to $\lesssim0.02$~pc per pixel for radial distances $\lesssim2$~kpc, so star-forming clumps in the cloud would be resolved if the cloud was in the Galactic disk. 
In addition, the high line ratios argue against ionization from young stars, as discussed in Section~\ref{sec:ionisation}. 

Based on the shell's morphology, velocity, line ratios and stellar associations, we thus conclude that the shell is most likely associated with the M81 group rather than being an interstellar or high galactic latitude cloud in the Galaxy.

\subsection{Is the gas tidally stripped from M81 or M82?}

The M81 group of galaxies began interacting about 1 Gyr ago, with the closest encounter occurring about 0.3 Gyr ago, during which gas was tidally stripped from the galaxies and redistributed across the field of the group \citep[e.g.][and references therein]{yun94,mayy06}.
If the shell was produced during the interactions, one would expect the spatial and dynamic properties of the shell to match those of the HI gas tidally stripped from the galaxies \citep[see, e.g., simulations of][]{yun94}. Inspection of the channel maps from \citet[][their Figure 2]{debl18} shows that there is tidally stripped HI gas with velocities ranging from $-260$ to 320 km/s, but the spatial distribution of the gas greatly varies across the range of velocities. HI gas with radial velocities below 20 km/s are only located at the position of M81 and southward, with no gas north of declination~$\approx60^{\circ}$.\edit1{ The velocity of the HI gas at the same declination as} the shell \edit1{($\delta\approx70^{\circ}$) is $\gtrapprox$160 km/s}, which is inconsistent with our derived shell velocity of $\approx-35$ km/s \citep{debl18,yun94,chyn08}.
It therefore seems unlikely that the \ha~shell is an ionized extension of the HI gas that was tidally stripped, implying either a separate stripping event \edit1{or a different mechanism for the origin of the shell. A separate tidal stripping event seems} unlikely as that would be expected to produce HI in addition to HII.
Further simulations of the M81 group interactions going back further in the history of the group (e.g. $\gtrsim1$ Gyr) are required in order to determine whether a separate tidal event have formed the \ha~shell.

\subsection{Is the shell lifted gas from the disk of M82?}


Could the shell be composed of gas that was lifted from the disk of M82 by the starburst, as was thought to have happened to create the \ha~cap \citep[e.g.,][]{devi99}? 
If the gas was pushed by superwinds from the starburst, it would take $\sim$50 Myr for the gas to reach its current location, assuming a wind velocity of $\approx800$ km/s on average \citep{lehn99}.
The epochs of peak starburst activity were modeled by \citet[][]{fors03} to have occurred $\sim10$ Myr ago followed by a second starburst $\sim5$ Myr ago. 
If gas was lifted from the disk of M82 to the current position of the shell, an additional starburst $\gtrsim50$~Myr ago is needed assuming an average wind velocity of $\lesssim800$ km/s. This event is not predicted by models of star formation in M82, which model star formation in M82 up to $\sim$100 Myr ago \citep[][and references therein]{fors03,yao09}. \edit1{It is possible that the tidal interations of M81 and M82 about 250 Myr ago produced a starburst in M82, which provided the energy to blow out the gas to the location of the shell.}
The duration of the starbursts are expected to be a few million years each \citep{fors03} so assuming a supernova energy production rate of $\sim10^{43}$ erg s$^{-1}$ \citep{chev85}, the total energy available ($\sim10^{56}$ erg) would be more than that required to lift the mass of the shell from the disk of M82 to its current position ($\lesssim10^{54}$ erg).
\edit1{This activity is not merely gravitational but also ballistic, requiring enough force against the ambient medium to raise the gas to the shell position. We carry out a back-of-the-envelope calculation to determine whether a blastwave could move the gas through ambient medium of density $n_H\approx0.01$ cm$^{-3}$. This number density corresponds to the average column density in the M81 group at a few tens of kpc from the galaxies in the HI measurements of \citet{debl18} and assuming a line-of-sight width of 5 kpc. The resulting Sedov-Taylor solution \citep[e.g.,][]{true99} requires an energy input from the M82 superwind according to the following relation: $L \approx 2~\times~10^{41}~\mathrm{erg~s}^{-1} (v/70~\mathrm{km~s}^{-1})^3 (R/44~\mathrm{kpc})^2 (n/0.01~\mathrm{cm}^{-3})$.
This energy can be supplied by the supernova energy production rate in M82 \citep[$\sim10^{43}$ erg s$^{-1}$; ][]{chev85}. 
While basic energetics arguments allow the possibility of the gas being blown out to the radius of the M82 shell, the question of whether the gas would survive within the hot wind from M82 remains.}

\edit1{Kelvin-Helmholtz and Rayleigh-Taylor instabilities have been shown to shred cold gas clouds entrained in winds within a timescale of $\sim1$~Myr \citep[e.g.,][]{coop09}
but recent analytic work and simulations have determined potentially viable mechanisms for the stable production and maintenance of cold gas in a hot wind  \citep[e.g.,][]{scan15,thom16,gron18,gron21,spar19,schn20,fiel21}.
In particular, recent simulations have shown that cold gas may be produced in the mixing layer between the cold cloud and hot (laminar) wind \citep[][]{gron18} as long as the cold clouds are large enough \citep[e.g., $R_{\mathrm{cloud}}\gtrsim 1-150$ pc;][]{gron18,spar19}.}
\edit1{Taking into account a filling factor of $10^{-4}$, individual ``cloudlets'' within the shell would take up a total volume of $\sim500^3$ pc$^3$. This size is above the minimum needed for survival predicted by these simulations, but the intial number of cloudlets would play an important role in determining the actual size and survivability. Cool clouds are prone to fragmentation \citep[e.g.,][]{mcco18,spar19} and large clouds in a turbulent hot medium are predicted to be shredded into many ``droplets'' spread out over a large area \citep[][]{gron21}. If this is the case for the shell, perhaps the shell started out with a smaller volume and was dispersed in a turbulent wind.
Further studies with longer timescales and larger box sizes, and with additional physics (such as magnetic fields, cosmic rays, etc.) are required to confirm whether this could be the process that created the \ha~shell.}

\edit1{In addition to gas being lifted from the disk, it is possible the superwind swept up existing gas surrounding M82. Another potential scenario is that the northern surroundings of M82 may also have originally been filled with tidally disrupted HI gas at similar levels to the HI observed in the southeast, and the M82 superwind ionized and pushed the tidally disrupted gas out to the location of the \ha~shell. The dynamics of the shell would require the superwind to have pushed the gas along the line-of-sight towards the observer, implying in the past there was a stronger radial component to the superwind than has been measured \citep[e.g.,][]{shop98}. In addition to more detailed simulations of the tidal disruption, further comparisons between the metallicity properties of the gas in the shell and the tidally disrupted HI gas would be helpful to determine whether they have the same origin.}

\subsection{Is the gas infalling from the intergalactic medium?}

If the gas in the shell is infalling from the intergalactic medium (IGM), the shell would need to have dynamics and metallicity consistent with an inflow. 
The \ha~shell has a relative radial velocity $\lesssim20$~km/s with respect to M81 \edit1{where the systemic velocity of M81 is $v_{\mathrm{M81}}\approx-15$~km/s \citep[][]{yun94}. The escape velocity of M81 at the projected distance of the shell is $v_{\mathrm{esc,M81}}\approx250$~km/s so the shell is} likely bound to the group and could be falling into its center-of-mass.
The shell has a large relative velocity with respect to M82 ($\approx250$~km/s) \edit1{which is greater than the escape velocity of M82 ($v_{\mathrm{esc,M82}}\approx50$~km/s)}. Consequently, the mass of M82 alone would not gravitationally capture the shell. If the shell is infalling, the gas likely will end up either as part of the intragroup medium or accrete onto M81 rather than M82.

The IGM is expected to have \edit1{on average} a metallicity of $\sim0.1$~Z$_{\odot}$ in the local Universe \edit1{though there is large variation of metallicities predicted in simulations due to poor mixing of the IGM \citep[e.g., a spread of metallicity of Z$_{\mathrm{IGM}}$/Z$_{\odot}\sim0.001-1$;][]{shul12}}. 
If the gas is infalling from the IGM for the first time, its metallicity should \edit1{fall within this range.}
As the \nii/\ha~ratio in the shell is likely boosted by shock emission, it is difficult to accurately determine an estimate of the metallicity in the gas.
\edit1{The \nii/\ha~ratios in the shell are consistent with \nii/\ha~ratios in the MAPPINGS III shock models for metallicities of $\sim0.2-2$~Z$_{\odot}$ \citep[see, e.g., Fig. 21 from][and discussion in Section~\ref{sec:ionisation}]{alle08}.}
Additional line information would help to further constrain the metallicity estimate.


As the position of the shell is well within the virial radius of the M81 group (at about $\frac{1}{2}\mathrm{R}_{vir}$ of M81; see Fig.~\ref{fig:halphaboth}), one potential scenario could be that intragroup gas was originally pressure supported by gas in the group, then the M82 superwinds removed gas along the northern minor axis of M82, which allowed the gas to free fall inwards and shock at a position closer into the galaxies within the virial radius of the group. 
Accretion shocks at or within the virial radius are expected to occur for halos of the sizes of the M81 group \citep[$\sim10^{12}$ M$_{\odot}$;][]{kara06,birn03}.
Thus the infalling gas may have experienced an accretion shock instead of (or in addition to) being shocked by the M82 superwind. 





\subsection{Composite origin scenario}



The final scenario we consider is one in which the M82 superwind entrained hot gas from the M82 disk and raised it to the position of the shell where it shocked and mixed with the intragroup gas, ionising the in situ gas and producing the \ha~emission we observe. 
Along with introducing density perturbations that seeded the precipitation of the circumgalactic gas \citep[as predicted in simulations, e.g.,][]{esme21},
the entrained gas in the superwind could have mixed with the in situ intragroup gas and enriched the gas. The low velocity dispersion from the line widths of the spectral data belies this theory as one would expect the mixing of the windblown gas with the intragroup gas to introduce turbulence into the medium. As such, perhaps the shell was enriched so far in the past that other relics of that event have long since vanished - such as through an early tidal event or through extended superwind events. More modeling is required to confirm or \edit1{disprove} these possibilities.

\section{Comparison with similar objects from literature}

Similar structures have been reported previously, such as ``Hanny's Voorwerp'' \edit1{near the spiral galaxy IC 2497} \citep[][]{lint09} and a cloud in the halo of M51 \citep{watk18}. These two gaseous structures were also discovered in \ha~emission and are notably closer \edit1{in projection} to their associated galaxies with projected distances of 25 kpc and 32 kpc, respectively. These distances place the clouds well within the virial radii of their host galaxies \edit1{whereas} the M82 shell is at the virial radius of M82 \edit1{\citep[i.e., the outer edge of the CGM;][]{tuml2017}, making it likely to be an intragroup cloud}.

In addition, many \ha~emitting objects have been found with close association to galaxies \citep[][and references therein]{bait19}.
Most of these structures appear to have tidal origins and many have an optical counterpart. We do not detect a continuum emission counterpart to the \ha~shell, though the presence of significant Galactic cirrus in the field makes the detection difficult and might mask the presence of a low surface brightness stellar counterpart to the shell. In the red giant branch stellar density maps of \citet[][]{smer20}, the whole field presented here is filled with stars. Inspection of their maps reveals no statistically significant overdensity of RGB stars at the location of the \ha~shell. In any case, no known or candidate dwarf galaxies are coincident with the shell location \citep[][]{chib09,okam19}. Deep followup is required to determine whether a significant stellar counterpart exists. If a significant metal-poor stellar counterpart were found it would suggest that the shell could be a dispossessed ``old disk'' component of M82, which \citet[][]{sofu98} theorize may have been ripped off of M82 during its tidal interactions $\sim1$ Gyr ago and has not yet been observed. Though ionization and origin mechanisms vary between these \ha~emitting clouds, the relative velocities of the \ha~structures and their associated galaxies range from 150 $-$ 400 km/s \citep{bait19}, which is consistent with the velocity difference observed between M82 and the \ha~shell and an M81 group association for the shell.

\section{Summary \& Conclusion} \label{sec:conclusion}

Deep \ha~imaging of the M81 group of galaxies with a novel upgrade to the Dragonfly Telephoto Array, a pathfinder version of the \dflong, has revealed a host of low surface brightness gaseous structures within the group. One significant structure is a colossal \ha-emitting ``shell'' of gas, over 0.8$^{\circ}$ in length, $\approx40$ kpc from the M82 galaxy in projection along its minor axis, i.e., at and potentially beyond the virial radius of the galaxy.
We argue that the shell is part of the M81 group of galaxies due to its morphology, velocity, and potential sources of ionisation (rather \edit1{than} being a chance projection of Galactic gas with the M81 group). In order to ionise the gas seen in the shell, an additional source of ionisation to the global UVB is required. We show that the shell is consistent with being shock ionised, either through incident superwinds from the M82 starburst or through accretion shocks as the gas is falling into the group of galaxies.
While the gas in the shell could have a tidal origin, the shell does not have similar velocity or spatial overlap with the extensive tidally stripped HI gas in the group, which one would expect to see if the gas was tidally stripped. We consider whether the gas could be M82 disk gas entrained in the superwind produced by the M82 starburst \edit1{or tidally stripped gas that was caught up in the superwind, but the question remains whether the cold gas would be expected to survive long enough to reach the location of the shell.}
Alternatively, the shell could have an external origin, with gas falling in from the intergalactic or intragroup medium. \edit1{With a radial velocity much less than the escape velocity of M81, the shell is likely bound to the group.} Further analysis of the M82 \ha~shell, including X-ray, UV, and/or deep visible spectral observations to pinpoint the ionisation source of the gas, is required to determine its origin.

Imaging of the M81 group of galaxies with the pathfinder \dfshort~serves as a test case for deep wide-field \ha~imaging of large-scale gaseous structures around nearby galaxies. This work foreshadows investigations of other nearby galaxies with \edit1{an upcoming} 120-lens \dflong~that will begin taking data in 2022. The \dfshort~upgrade is based on the pathfinder which was used to collect the data presented here (further details on the \dfshort~ instrument are described in Chen et al., in prep). \dfshort~\edit1{will have $40\times$ the collecting area of the pathfinder and will reach the limits presented here in under one hour.}

\acknowledgments
This work was benefited greatly from conversation and correspondence with Crystal Martin, Joss Bland-Hawthorn, Chris Matzner, and Natasha F\"{o}rster-Schreiber, as well as through useful feedback from the anonymous reviewer. We thank you for your comments and guidance. We are very grateful to the staff at New Mexico Skies Observatories, without whom this work couldn't have been carried out.
The work done by D.M.L. for this paper was supported through grants from 
the Ontario government. We are thankful for contributions from the Dunlap Institute (funded through an endowment established by the David Dunlap family and the University of Toronto), the Natural Sciences and Engineering Research Council of Canada (NSERC), and the National Science Foundation (NSF), without which this research would not have been possible.
S.D. is supported by NASA through Hubble Fellowship grant HST-HF2-51454.001- A awarded by the Space Telescope Science Institute, which is operated by the Association of Universities for Research in Astronomy, Incorporated, under NASA contract NAS5-26555.

%

\facilities{Dragonfly Telephoto Array; pathfinder \dflong; Keck LRIS}


\software{astropy \citep{astropy},  
          source-extractor \citep{sourceextractor},
          pypeit \citep{pypeit:zenodo},
          iraf \citep[][]{tody86}
          }\\\\



\vspace{10mm}

\appendix
\section{\dfshort~Data Reduction with DFReduce}\label{appendix:datared}
The data reduction pipeline for Dragonfly data is described in detail in \citet{dani20} and \citet{zhan21j}, and we refer the interested reader to these publications for details on the data reduction procedures. In this \edit1{Appendix}, we summarize the basic steps and any differing procedures for narrowband data reduction. The narrowband and broadband data were reduced using the \texttt{DFReduce} package (Greco et al., in prep.). While the reduction procedure for both datasets was similar, they were reduced separately due to the differing pixel scales in the CCD cameras used by the 48-lens Dragonfly and the pathfinder \dfshort.

The images were dark-subtracted and flat-fielded. For the narrowband data, flat-fielding was carried out using master flats created from flats at the same tilt and pointing as the data. 
After dark subtraction and flat-fielding, the data was passed through a series of image quality checks to throw out `bad' data frames. These checks included limits on the FWHM, ellipticity and number of point sources, as well as removing frames which are determined to be off-target by more than 1.5 degrees.

At this point, the frames went through the first round of sky subtraction. The background sky in each frame was separately modeled with a 3$^{\mathrm{rd}}$ order polynomial and subtracted from each image. Sky subtraction for emission lines is often a large source of error due to the relative strength of sky lines to astrophysical emission lines from diffuse gas, and by masking and fitting a low order polynomial across each image, we removed large scale emission on scales of $\sim$0.5 deg. 
The frames were then registered to align them onto the same grid.
The average magnitude zeropoint level of the point sources in each frame was then calculated and frames with a difference in zeropoint of greater than 0.1 mag from the median zeropoint of all frames taken by the same camera were rejected. For both the narrowband and the $r$ broadband data, the zeropoint level was calculated by comparing the magnitude in the data frames to the $r$-band magnitude from the The AAVSO Photometric All-Sky Survey catalogue. This proved suitable for rejecting non-photometric frames for all the data.

After rejection of bad frames, the images were stacked together taking the median of each pixel value to create median coadds separately for the narrowband and broadband data.
All the accepted frames then went through another round of data reduction, repeating all the steps above up to registering the frame, but with one change: during the sky subtraction, the median coadd was used to create a mask for all point sources to create the sky model to better subtract the sky from the data.
After a final registration, the data was combined into average stacks to form final science images in H$\alpha$, \textsc{[Nii]}, $g$, and $r$, with a common pixel scale of $2.1''$ per pixel.




\bibliography{references}{}
\bibliographystyle{aasjournal}



\end{document}